\documentclass[a4paper]{article}  
\usepackage[a4paper,margin=0.75in]{geometry}
\usepackage[affil-it]{authblk}
\pdfoutput=1

\usepackage{siunitx,amsmath,amssymb,graphicx,color}
\usepackage[version=4]{mhchem}
\graphicspath{ {figures/} }

\usepackage{xr}
\usepackage{hyperref}

\usepackage{setspace}
\usepackage[normalem]{ulem}

\title{Painting with bacteria: Smart templated self assembly using motile bacteria}

\author[1]{Jochen Arlt%
  \thanks{Electronic address: \texttt{j.arlt@ed.ac.uk}; Corresponding author}}
\author[1]{Vincent A Martinez}
\author[1]{Angela Dawson}
\author[2]{Teuta Pilizota}
\author[1]{Wilson C K Poon}

\affil[1]{School of Physics \& Astronomy, The University of Edinburgh, Peter Guthrie Tait Road, Edinburgh EH9 3FD, United Kingdom}
\affil[2]{School of Biological Sciences and Centre for Synthetic and Systems Biology, The University of Edinburgh, Roger Land Building, Alexander Crum Brown Road, Edinburgh EH9 3FF, UK}

\date{}

\begin{document}

\maketitle


\begin{abstract}
External control of the swimming speed of `active particles' can be used to self assemble designer structures in situ on the $\si{\micro\meter}$ to $\si{\milli\meter}$ scale. We demonstrate such reconfigurable templated active self assembly in a fluid environment using light powered strains of {\it Escherichia coli}. The physics and biology controlling the sharpness and formation speed of patterns is investigated using a bespoke fast-responding strain.
\end{abstract}

Micro- and nano-fabrication can revolutionise many areas of technology, including personalised medicine. There are two conceptually distinct ways to construct structures on the $\SI{10}{\nano\meter}$ to \SI{10}{\micro\meter} scale: lithography, which uses `scalpels' such as chemical etching or electron beams, or self assembly\cite{Whitesides2002}, in which microscopic `Lego components' move themselves into position. Both equilibrium phase transitions (e.g.~crystallization) and non-equilibrium processes are exploited for self assembly. In either case, external templates can be used to direct the process, with reconfigurable templates offering programmability. Self assembly was originally inspired by chemistry and biology, where the components are individual molecules. Increasingly, colloidal building blocks are used, with bespoke particle shape, size and interaction, e.g.~`patchy particles' with heterogeneous surface chemistry\cite{Sciortino2011}. Active, or self-propelled, colloids open up further opportunities. We show how to assemble structures on the \si{\micro\meter} to \si{\milli\meter} scale that are reconfigurable in real time using \textit{Escherichia coli} bacteria -- `living active colloids' -- that swim only when illuminated\cite{Walter2007}. The process is directed by a smart, or programmable, external template applied by a spatial light modulator. 

Active colloids, or self-propelled micro-swimmers, are attracting significant recent attention\cite{poon2013physics} as `active matter'\cite{Sriram,Joanny2014}. 
They violate time-reversal symmetry\cite{CatesReview2012}, and may be used, e.g.,  to transport colloidal `cargos'\cite{SenCargo}. 
For both fundamental physics and applications, external control of swimming, e.g.~using particles with light-activated self-propulsion\cite{Bechinger2016,Li2016,SenSilver,PalacciCrystal}, opens up many new possibilities. 
Thus, e.g., light-activated motile bacteria can be used to actuate and control micro-machinery \cite{Leonardo2017}. 

The self assembly of micro-swimmers into clusters of tens of particles has already been demonstrated\cite{Jana,BocquetCluster,PalacciCrystal}. Recent simulations\cite{Cates2016} suggest that the patterned illumination of light-activated swimmers can be used for the templated self assembly\cite{Whitesides2002} of designer structures comprising $10^3$-$10^4$ particles. Real-time reconfiguration of the light field then allows {\it smart} templated active self assembly (STASA), which we here implement for the first time using light-controlled motile bacteria. 

{\it E. coli} bacteria\cite{Poon2016a} (cell body $\approx \SI{2}{\micro\meter} \times \SI{1}{\micro\meter}$) swim by turning $\approx 7$-$\SI{10}{\micro\meter}$ long helical flagella using membrane-embedded rotary motors powered by a protonmotive force (PMF) that arises from active pumping  of \ce{H+} to the extracellular medium\cite{Berg1977}.  
Unlike all synthetic active colloids to date and most bacteria, {\it E. coli} can generate PMF in nutrient-free motility buffer\cite{AdlerEnviron} by utilising internal resources and oxygen (\ce{O2}) to produce energetic electron pairs. 
These release their energy stepwise along an electrochemical potential ladder of respiratory enzymes located in the inner cell membrane, generating a PMF of $\approx \SI{-150}{\milli\volt}$. The electron pair ultimately passes to and reduces \ce{O2} to water. 
Thus, with no \ce{O2}, PMF = 0 and swimming ceases\cite{Poon2016a}. If cells under anaerobic conditions can express proteorhodopsin (PR)\cite{Walter2007}, a green-photon-driven proton pump\cite{Bela2000}, then they will swim only when suitably illuminated: these are living analogues of synthetic light-activated colloidal swimmers.\cite{Bechinger2016,Li2016,SenSilver,PalacciCrystal} 

We show below that the speed with which such cells respond to changes in illumination is crucial for successful bacterial STASA. For this work, we constructed a PR-bearing mutant (AD10) that stops much faster when illumination ceases than previously-reported\cite{Walter2007,Armitage2013}, by deleting the {\it unc} gene cluster\cite{unc1980} encoding the $\textrm F_{1} \textrm F_{\textrm o}$-ATPase membrane protein complex (see SI \S\ref{sec:SI1}), so that these enzymes cannot act in reverse in darkness to continue to export protons and sustain a PMF\cite{Keis2006}.

We suspended cells in phosphate motility buffer at optical density OD~$\lesssim 8$ (cell-body volume fraction $\approx 1.1\%$)\cite{Poon2016a} and sealed $\SI{2}{\micro\liter}$ into \SI{20}{\micro\meter} high flat capillaries, where cells swim in two dimensions but have enough room to `overtake' each other in all three spatial dimensions. Differential dynamic microscopy (DDM)\cite{WilsonDDM} returned an averaged speed $\bar v \approx \SI{30}{\micro\meter\per\second}$ and $\beta \approx 20\%$ of non-motile organisms  at OD=1 under fully-oxygenated conditions. (Note that `non-motile' = cells that can never swim;  `stationary' = non-swimming cells capable of motility when illuminated.)

\begin{figure}
\begin{center}
\includegraphics[width=\columnwidth]{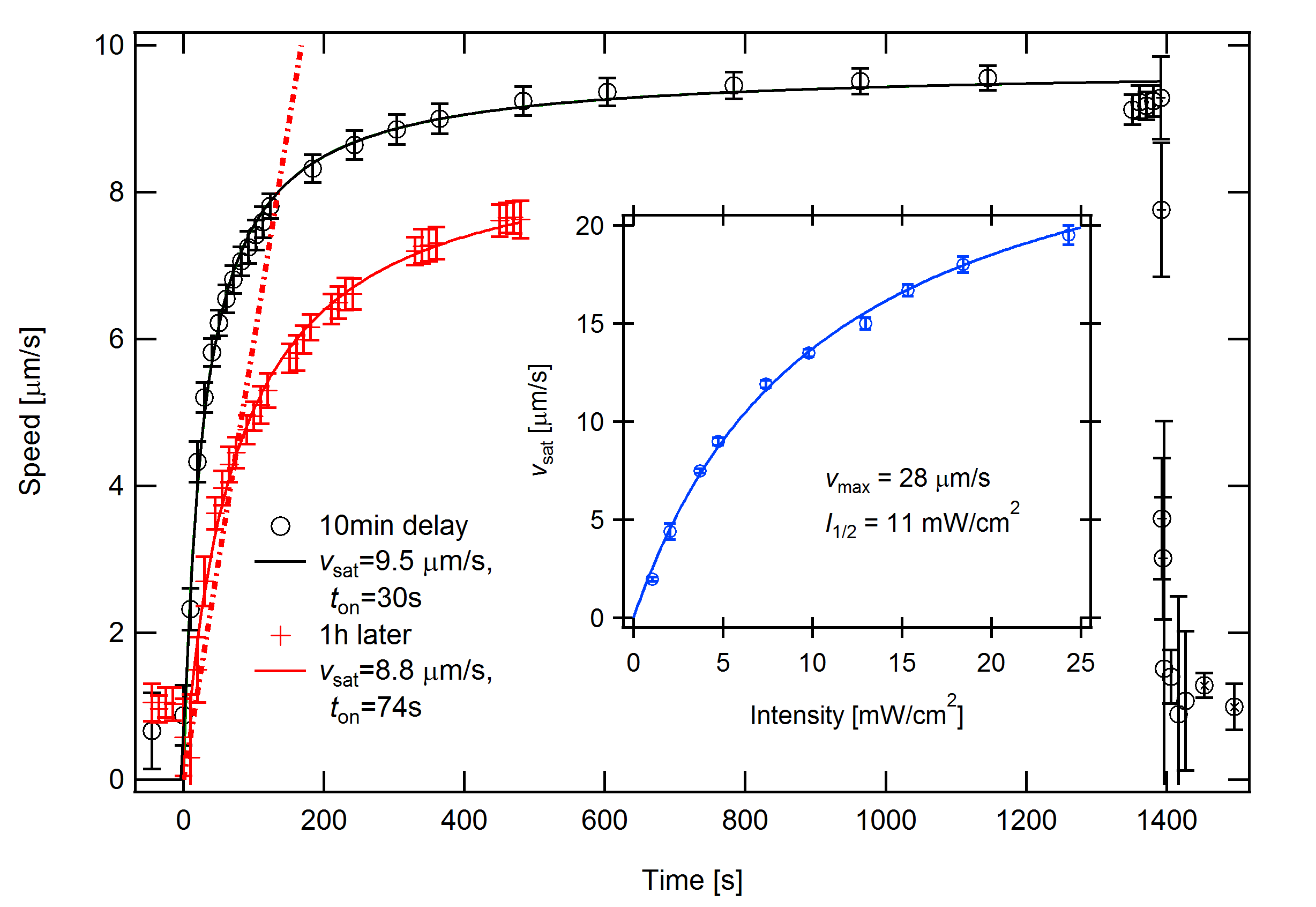}
\caption{Population-averaged speed {\it vs.}~time for stationary {\it E. coli} AD10, light on at $t = 0$ and off at $t \approx \SI{1400}{\second}$ ($\circ$). Data points just after switch-off are $\SI{1}{\second}$ apart to resolve the fast stopping. This sample was measured soon after the cells had run out of \ce{O2}. Cells depleted of  \ce{O2} for $\SI{1}{\hour}$ longer accelerated less rapidly (${\color{red} \boldsymbol{+}}$). Full lines: fits to $\bar v(t) = \bar v_{\rm sat} t/(t+\tau_{\rm on})$. The dot-dashed line has slope \SI{0.05}{\micro\meter\per\square\second}, the fitted acceleration from `box emptying' experiments starting from stationary cells. Inset: dependence of $\bar v_{\rm sat}$ on incident light intensity together with fit to $\bar v_{\rm sat}(\mathcal{I}) = \bar v_{\rm max} \mathcal{I}/(\mathcal{I}+\mathcal{I}_{1/2})$ (see SI \S\ref{sec:SI3}).} 
\label{fig:speed}
\end{center}
\end{figure}

Motile cells were allowed to swim until \ce{O2} was depleted and $\bar v$ dropped abruptly to zero after a few minutes\cite{Poon2016a} (see SI \S\ref{sec:SI2} and Fig.~\ref{fig:SI:StartupAndIntensity}(a)). 
After these cells were left in the dark for $\approx \SI{10}{\minute}$, green illumination was turned on (510 -- $\SI{560}{\nano\meter}$, intensity $\mathcal{I} \approx\SI{5}{\milli\watt\per\centi\meter\squared}$ at the sample). The stationary cells accelerated uniformly before saturating, Fig.~\ref{fig:speed}.
Fitting the data to $\bar v(t) =\bar v_{\rm sat} t/(t+\tau_{\rm on})$ gives $\bar v_{\rm sat} = \SI{9.5}{\micro\meter\per\second}$, $\tau_{\rm on} = \SI{30}{\second}$.  
When illumination ceased, $\bar v$ dropped within $\tau_{\rm off} \lesssim \SI{1}{\second}$, but never quite to zero -- it is unclear why a few cells ($< 1\%$) continued to swim. $\bar v_{\rm sat}$ increased with $\mathcal{I}$, Fig.~\ref{fig:speed}~(inset), up to $\lesssim \SI{27}{\micro\meter\per\second}$. $\{ \bar v, \beta, v_{\rm sat}, \tau_{\rm on}$\} changed over hours as cells aged. 

The discharging of the PMF through the membrane (capacitance $C \gtrsim 10^{-14}\, \si{\farad}$) and rotary motors (total resistance $R \lesssim 10^{14}\,\si{\ohm}$) upon cessation of illumination should take $RC \sim \SI{1}{\second}$ (see SI \S\ref{sec:SI3} for details), which explains the observed $\tau_{\rm off}$. Consistent with this interpretation, $\tau_{\rm off}$ is approximately independent of the starting speed of decelerating cells (see SI \S\ref{sec:SI2} and Fig.~\ref{fig:SI:StoppingVsSpeed}(a)).  The observed $\tau_{\rm on} \approx \SI{30}{\second}$ is likely controlled by the rate constant\cite{Leake2006} for stator units to come on and off motors, $k_{\rm stator} \approx \SI{0.04}{\per\second} \sim \tau_{\rm on}^{-1}$. In sustained darkness, motors disassemble in PR-bearing {\it E. coli}, and full `motor resurrection' upon illumination takes\cite{Armitage2013} $\sim \SI{200}{\second}$, in agreement with our data, Fig.~\ref{fig:speed}. Consistent with this, if illumination is lowered and then restored rapidly (i.e.~in $\ll k_{\rm stator}^{-1}$), motors do not have time to disassemble and $\tau_{\rm on} \sim RC$ (see SI \S\ref{sec:SI2} and Fig.~\ref{fig:SI:StoppingVsSpeed}(b)).

\begin{figure}
\begin{center}
\includegraphics[width=0.99\textwidth]{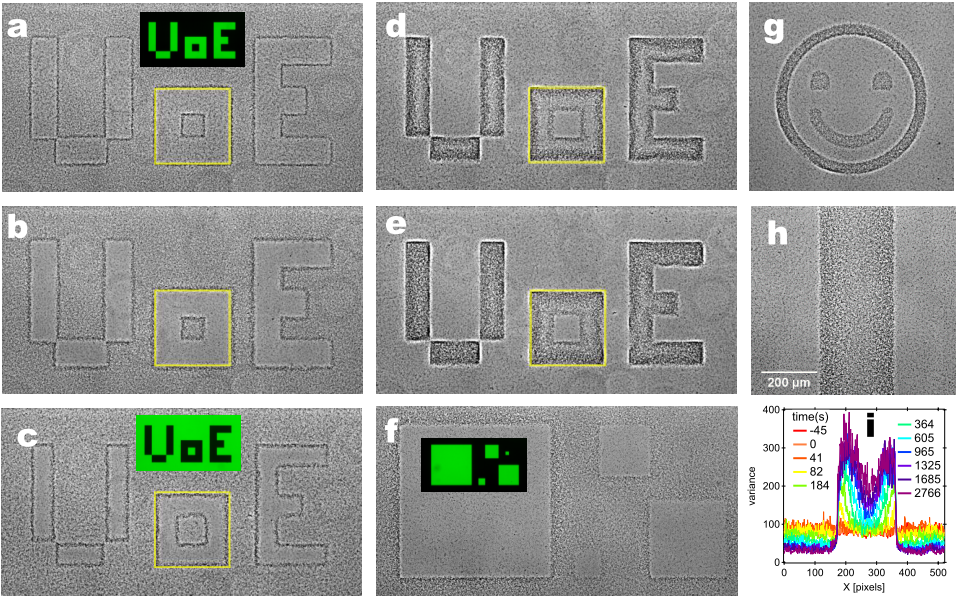}
\caption{Positive and negative masks projected onto initially stationary cells: (a) sample illuminated for \SI{1}{\minute} with a positive `UoE' (University of Edinburgh) pattern shown in inset,
 (b) same after \SI{9}{\minute} of illumination, (c) \SI{1}{\minute} after switching to the negative pattern shown in inset (with green = bright), as well as (d) \SI{6}{\minute}  and (e) \SI{12}{\minute} after switching. Yellow square = boundary between light and dark regions of the `o'. (f) Samples illuminated with pattern of squares shown in inset. (g) Negative smiley pattern. (h) Negative strip together with (i) time evolution of the strip's density profile. Scale bar in (h) applies throughout.}
\label{fig:patterns}
\end{center}



\end{figure}

Our STASA protocol uses a spatial light modulator (SLM) to project patterns of illumination onto a uniform field of cells rendered stationary by \ce{O2} exhaustion. 
We first projected a positive mask with bright features on a dark background, Fig.~\ref{fig:patterns}a (inset). 
The width of the features, \SI{90}{\micro\meter}, is comparable to the persistence length (i.e. the length scale over which the cell's motion can be approximated as a straight line) of smooth-swimming {\it E.~coli}\cite{Wu2006}. 
Over time, the mask pattern was replicated in the cell population with increasing clarity: cells swim out of illuminated areas, stop within $\tau_{\rm off} \sim \SI{1}{\second}$, and accumulate at the edges just {\it outside} the boundary of these areas, Fig.~\ref{fig:patterns}b (also see \href{run:./SI-Movie1-UoE-startup-2fps-5xSpeedup-Time-uncomp.avi}{SI movie~\ref{mov:SI:startup}}). 
Next we abruptly switched to a negative mask (dark pattern against bright background), Fig.~\ref{fig:patterns}c (inset). Edge-accumulated cells start swimming, about half of which arrive inside the dark features immediately and stop within $\tau_{\rm off}$, rapidly outlining the {\it inside} edge of the masked areas, Fig.~\ref{fig:patterns}c. 
Thereafter, the dark areas progressively filled up: bacteria continually arrive from the outside, which acts as an essentially infinite reservoir of swimming cells, Fig.~\ref{fig:patterns}d-e (also see \href{run:./SI-Movie2-UoE-inversion-2fps-5xSpeedup-Time-uncomp.avi}{SI~movies~\ref{mov:SI:inversion}} and \href{run:./SI-Movie3-UoE-TimeLapse.avi}{\ref{mov:SI:timelapse}}). 
Other patterns can be `painted' using this method, Fig.~\ref{fig:patterns}g.
In contrast to static `bacterial painting' recently demonstrated~\cite{Fernandez-Rodriguez2017}, our method is dynamic and reconfigurable in real time within our liquid medium.

\begin{figure}
\begin{center}
\includegraphics[width=1\textwidth]{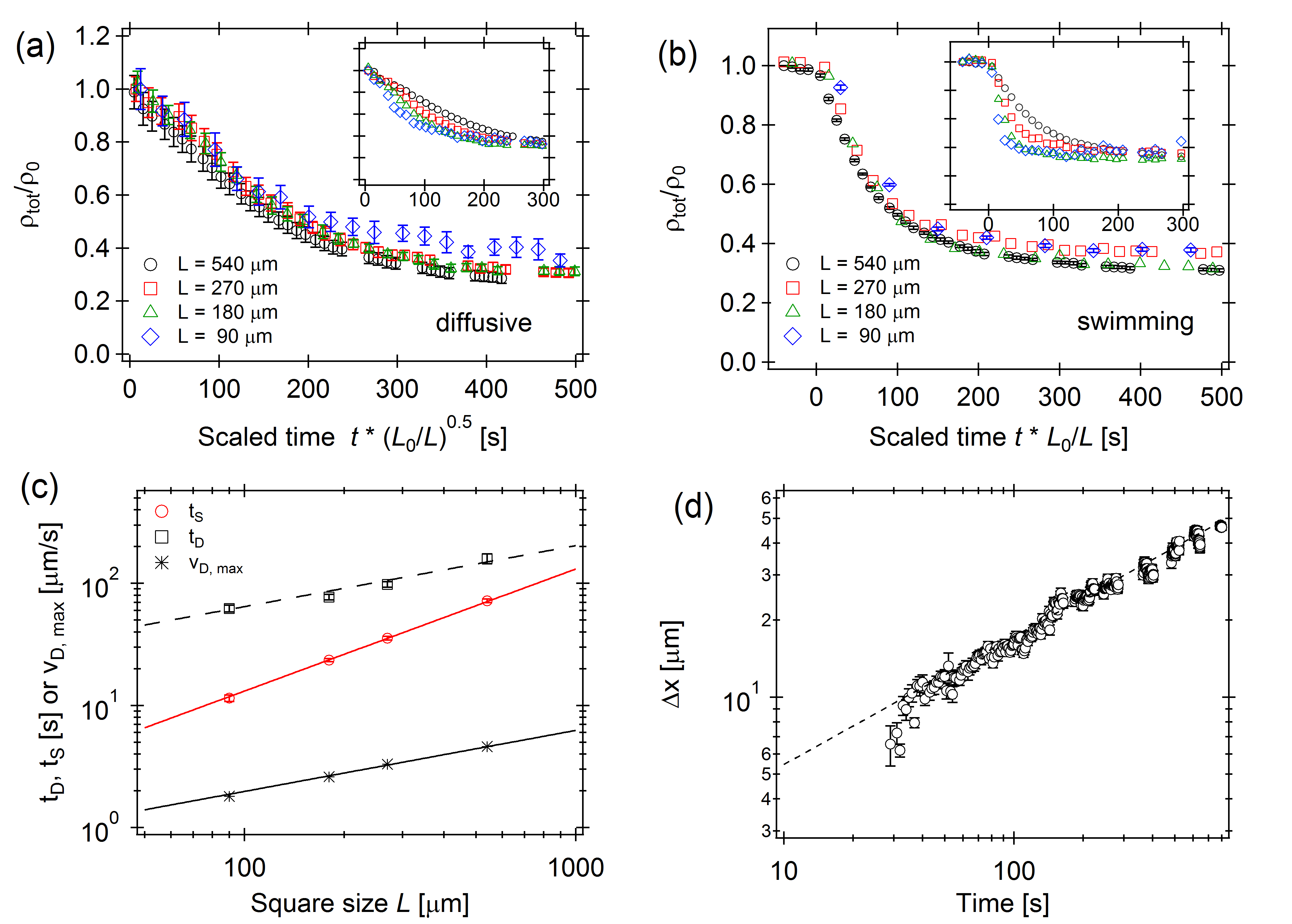}
\caption{Characterising the dynamics of pattern formation: Time evolution of normalised bacterial density (see SI \S\ref{sec:SI:DDM}) inside squares of different dimensions $L$ (inset,  Fig.~\ref{fig:patterns}f), starting from (a) diffusive state and (b) uniformly swimming state. Inset: data plotted against time; main plots: data collapse when scaling time by $\sqrt{L_0/L}$  in (a) and $L_0/L$ in (b) (both with $L_0=\SI{540}{\micro\meter}$)  (c) Scaling of `box emptying times' $\mathfrak{t}_{\rm D}$, $\mathfrak{t}_{\rm S}$ as well as $\bar v_{\rm D, max}$ (the maximum swimming speed reached when starting with diffusive cells) with square size $L$. (d) Shift of the peak position over time in the dark stripe, Fig.~\ref{fig:patterns}h; dashed line = power law fit of slope 0.5. }
\label{fig:scaling}
\end{center}
\end{figure}

To understand how fast such reconfigurations of spatial patterns can take place, we experimented with square masks, Fig.~\ref{fig:patterns}f, using two initial conditions: a (dark) field of stationary cells ($\bar v(0) = 0$, i.e., diffusive), and a field of (illuminated) cells that have reached speed saturation ($\bar v(0) = \bar v_{\rm sat}$). We studied the evolution of cell density (see Methods and SI \S\ref{sec:SI:DDM}) and speed inside bright squares against a dark background. Starting with diffusive cells, $\bar v(t) = \bar v_{\rm sat} t/(t + \tau_{\rm on}) \approx at$ with $a = \bar v_{\rm sat}/\tau_{\rm on}$ for $t \lesssim \tau_{\rm on}$. The time to empty an $L \times L$ square should scale as
\begin{equation}
\mathfrak{t}_{\rm D} (L) \sim (2L/a)^{0.5}. \label{eq:Tdiff} 
\end{equation}
The measured cell density evolution, $\rho_{\rm tot}(t)$, in illuminated squares of different sizes is shown in Fig.~\ref{fig:scaling}a, from which we extracted $\mathfrak{t}_{\rm D}$, the time for $\rho_{\rm tot}$ to drop to $1/e$ of its decay.  As expected, $\mathfrak{t}_{\rm D}(L) \sim L^{0.5}$, Fig.~\ref{fig:scaling}c.  Fitting Eqn.~(\ref{eq:Tdiff}) gives $a = \SI{0.05}{\micro\meter\per\square\second}$, which is a reasonable average acceleration over the first $\sim \SI{100}{\second}$ for the data in Fig.~\ref{fig:speed} (where the dot-dash line has slope \SI{0.05}{\micro\meter\per\square\second}). 
Interestingly, $\bar v_{\rm D, max}$, the maximum average swimming speed reached during emptying depends on the box size, displaying a $L^{0.5}$ scaling consistent with our model (fig.~\ref{fig:scaling}c, and SI \S\ref{sec:SI:squares}).

On the other hand, starting with cells all swimming (on average) at $\bar v_{\rm sat}$, a square should empty in
\begin{equation}
\mathfrak{t}_{\rm S} (L) \sim L/\bar v_{\rm sat}. \label{eq:Tswim}
\end{equation}
From the measured $\rho_{\rm tot}(t)$ in this case, Fig.~\ref{fig:scaling}b, we deduced a `box emptying time' and find, Fig.~\ref{fig:scaling}c, the expected scaling, $\mathfrak{t}_{\rm S}(L) \sim L$. 
Fitting Eqn.~(\ref{eq:Tswim}) gives $\bar v_{\rm sat} \approx \SI{8}{\micro\meter\per\second}$, consistent with the measured $\bar v_{\rm sat} = \SI{9.5}{\micro\meter\per\second}$. 
Equations~\ref{eq:Tdiff} and \ref{eq:Tswim} are also confirmed by the data collapse achieved for $\rho_{\rm tot}(t)$ when we scale the time axis by $L^{0.5}$ and $L$ respectively, Fig~\ref{fig:scaling}a,b. 

Further insight on the time scale for feature formation comes from studying cells moving into a dark area. We projected a $\SI{270}{\micro\meter}$-wide dark strip onto a field of 
initially diffusive cells, fig.~\ref{fig:patterns}h. 
An edge rapidly developed just inside the darkened area, corresponding to twin peaks in the spatial cell density profile, fig.~\ref{fig:patterns}i (as measured by the variance of image intensity, see SI \S\ref{sec:SI:Variance}), which broaden and fill the stripe over time.
The squared displacement of this peak shows diffusive dynamics, $\Delta x^2 = 2Dt$, Fig.~\ref{fig:scaling}d, with a fitted $D = \SI{0.74}{\square\micro\meter\per\second}$. 
The interface between bright and dark areas acts as an absorber of swimmers arriving with an effective diffusive process. Due to conservation of mass the diffusive propagation of the peak is thus expected, although quantitative aspects remain to be elucidated.


\begin{figure}
\begin{center}
\includegraphics[width=0.95\textwidth]{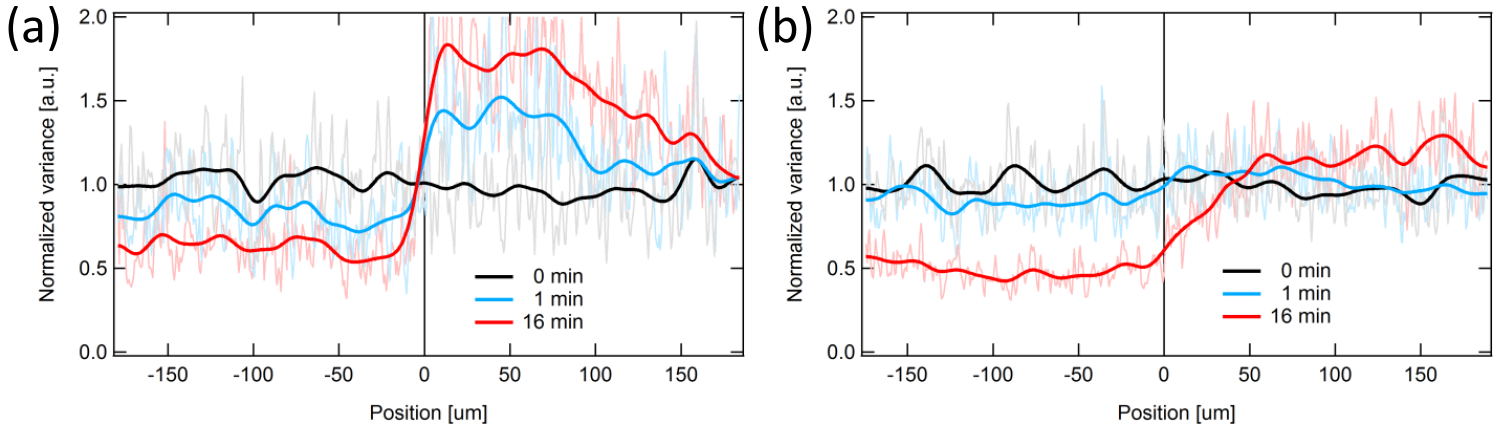}
\caption{Response to a step illumination pattern (left bright, right dark) projected at time $t = 0$ by two strains starting from initially uniformly swimming cells. Faint lines show the normalised variance (a measure of cell density) profile, while bold lines show numerically smoothed profiles. (a) AD10 with $\tau_{\rm off} \lesssim \SI{1}{\second}$ and (b) AD57 with $\tau_{\rm off} \gtrsim \SI{1}{\minute}$ (Corresponding images in SI fig.~\ref{fig:SI:edge}).}
\label{fig:shortOffTime}
\end{center}
\end{figure}

Finally, we seek to understand what controls the sharpness of the self-assembled features using our method. A stark demonstration of the answer comes from projecting a step-function pattern of light-dark illumination on initially uniformly swimming cells (see SI \S\ref{sec:SI:AD10vsAD57} for details). Figure~\ref{fig:shortOffTime} shows the resulting density of cells (from the intensity variance, see SI \S\ref{sec:SI:Variance}) for two strains, AD10, with $\tau_{\rm stop} \lesssim \SI{1}{\second}$, and AD57, a strain lacking the {\it unc} cluster deletion and having $\tau_{\rm off} > \SI{1}{\minute}$ (see Fig.~\ref{fig:SI:edge} for images). In the case of AD10, the illumination pattern is clearly replicated in cell density within \SI{1}{\minute}, and the steady state shows a sharp feature with half width $\lesssim \SI{10}{\micro\meter}$. For AD57, it takes $\gg \SI{1}{\minute}$ to establish a corresponding density pattern, which, in the steady state, is $\gtrsim \SI{50}{\micro\meter}$ wide. Thus, a short enough $\tau_{\rm off}$ is necessary for the fast establishment of sharp features using our STASA protocol.

In sum, we have demonstrated how to perform smart templated self assembly using a PR-bearing strain of {\it E. coli} with {\it unc} cluster deletion dispersed in anaerobic motility buffer. Spatial patterns of light are reproduced sharply in cell density, with the time scale of pattern formation controlled by the `on time' and the sharpness of the features controlled by the `off time'. This opens up many opportunities. Thus, e.g., it should now be possible to test a fundamental prediction that the product of density and speed is constant in a system with spatially-dependent activity\cite{TailleurEPL}. Practically, we can go beyond a recent demonstration\cite{Leonardo2017} of the control of lithographed micro-devices using light-activated bacteria: STASA allows the assembly and real-time reconfiguration of the devices themselves using the same bacteria\cite{Cates2016}. On the other hand, if cell density is used to trigger the secretion of adhesive biopolymers, bacterial STASA offers a route for manufacturing bespoke 10-\SI{100}{\micro\meter} microstructures, which remains a challenging length scale for current 3D printing.


\section*{Methods}
\label{sec:material}

\subsection*{Cells}
The {\it Escherichia~coli} used in this work (AD10 \& AD57) are smooth-swimming mutants of AB1157 transformed by a plasmid expressing SAR86 $\gamma$-proteorhodopsin (a gift from Jan Liphardt, UC Berkley), similar to the original study\cite{Walter2007}.
To achieve better light-switching response, we also deleted the $unc$ gene complex, thus removing the coupling between PMF and ATP\cite{unc1980}.

Overnight cultures were grown aerobically in $\SI{10}{\milli\liter}$ Luria-Bertani Broth (LB) using an orbital shaker at 30$^\circ$C and 200~rpm. A fresh culture was inoculated as 1:100 dilution of overnight-grown cells in $\SI{35}{\milli\liter}$ tryptone broth (TB) and grown for $\SI{4}{\hour}$ (when an optical density of OD$_{600}\approx0.2$ was reached). The production of proteorhodopsin (PR) was induced by adding arabinose to a concentration of $\SI{1}{\milli\mole}$ as well as the necessary cofactor all-trans-retinal to $\SI{10}{\micro\mole}$ in the growth medium.
Cells were incubated under the same conditions for a further hour to allow protein expression to take place and then transferred to motility buffer (MB, pH = 7.0, 6.2 mM K$_2$HPO$_4$, 3.8 mM KH$_2$PO$_4$, 67 mM NaCl and 0.1 mM EDTA).
Cells were re-suspended in MB after a single filtration (0.45~$\mu$m HATF filter, Millipore), yielding high-concentration stock solutions (OD $\approx 8$ to 10), which were then diluted with MB as required.

Suspensions were loaded into $\SI{2}{\micro\liter}$ sample chambers (SC-20-01-08-B, Leja, NL, $\SI{6}{\milli\meter} \times \SI{10}{\milli\meter} \times \SI{20}{\micro\meter}$), giving a quasi-2D environment for {\it E. coli} swimming. The chamber was sealed using vaseline to stop flow and oxygen supply, and placed onto a microscope for video recording. {\it E.~coli} initially swims under endogenous metabolism, which depletes the oxygen within the sample cells\cite{Poon2016a}. Once oxygen is depleted, the cells stop swimming unless illuminated by green light. 

\subsection*{Optical setup}
\label{sec:microscope}
The samples were observed using a Nikon TE2000 inverted microscope with a phase contrast objective (PF~10$\times$/0.3). Time series of $\approx \SI{40}{\second}$-movies (100 frames per second) were recorded using a fast CMOS camera (MC~1362, Mikrotron). A long-pass filter (RG630, Schott Glass) was introduced into the bright-field light path to ensure that the light required for imaging did not activate PR. The light controlling the swimming behavior of the bacteria was provided by an LED light source (Sola SE II, Lumencor) filtered to a green wavelength range ($510$ -- $\SI{560}{\nano\meter}$) to overlap with the absorption peak of PR\cite{Walter2007}.  
This light was reflected off a digital mirror device (DLP6500, Texas Instruments), which was imaged onto the microscope sample plane using a demagnifying relay telescope, leading to a resolution of $\SI{2.7}{\micro\meter}$ per pixel.
The light was coupled into the transmission path by introducing a dichroic mirror in between the long working distance phase contrast condenser and the sample.
This `trans' geometry allowed us to illuminate uniformly a circular area of $\approx\SI{2.9}{\milli\meter}$ in diameter, larger than the objective's field of view.
Computer control of the digital mirror device gave us precise spatial and temporal control of the illumination pattern.

\subsection*{Motility and density measurements.}
The motility of the samples was quantified using differential dynamic microscopy (DDM)\cite{CerbinoDDM1}, which yields the average swimming speed $\bar v$ and fraction of non-motile bacteria $\beta$ from low magnification movies\cite{WilsonDDM, MartinezDDM, Poon2016a}.
By comparing the amplitude of the DDM signal to that of a reference sample, it can also give the relative bacteria density $\rho/\rho_0$ (see SI \S\ref{sec:SI:DDM} for details). However, the spatial resolution of this method of density measurement is very limited.  In experiments where we need to measure density profiles with sharp spatial variations, we used instead the local variance of the image intensity to provide estimates of local cell density (see SI~\S\ref{sec:SI:Variance} for details).



The work is funded by an ERC Advanced Grant (ADG-PHYSAPS). We thank M.E. Cates and J. Walter for useful discussions.
 
\clearpage


\newcommand{\beginsupplement}{%
        \setcounter{table}{0}
        \renewcommand{\thetable}{S\arabic{table}}%
        \setcounter{figure}{0}
        \renewcommand{\thefigure}{S\arabic{figure}}%
     }

\beginsupplement
\section{Supplementary Information}

\subsection{Bacterial strains} \label{sec:SI1}

AD10 was constructed using our laboratory stock of {\it E.coli} AB1157~\cite{Dewitt1962}.    
All of the genes present within the operon encoding the ATP synthase complex ({\it atp}I, {\it atp}B, {\it atp}E, {\it atp}F, {\it atp}H, {\it atp}A,  {\it atp}G, {\it atp}D, {\it atp}C) were deleted in a single step by allelic replacement using a recombinant pTOF24 vector \cite{Merlin2002} containing 400bp homology arms flanking the deletion on each side.  
This resulted in the $\Delta unc$ strain AD10 with a deletion of 7504bp (corresponding to position 3915552-3923056 of the {\it E.coli} K12 MG1655 chromosome).   
P1 transduction was used to delete the {\it che}Y gene in both this strain and the parental strain (to generate AD57) using JW1871 (BW25113 {\it che}Y) as donor \cite{Baba2006}.  
Diagnostic PCR reactions were performed to confirm genotypes.  
Both strains were transformed with plasmid pBAD-His C (a kind gift from Jan Liphardt, UC Berkley) encoding the SAR86 $\gamma$-proteobacterial photorhodopsin variant \cite{Armitage2013}.

\subsection{Saturation, startup and stopping behaviour} \label{sec:SI2}

Here we give further information and detailed data supporting the claims made in the main text about the dependence of average swimming speed, $\bar v$, on light intensity ($\mathcal{I}$), startup time ($\tau_{\rm on}$), and stopping time ($\tau_{\rm off}$). 

\begin{figure}[bht]
\begin{center}
	\includegraphics[width=0.48\textwidth]{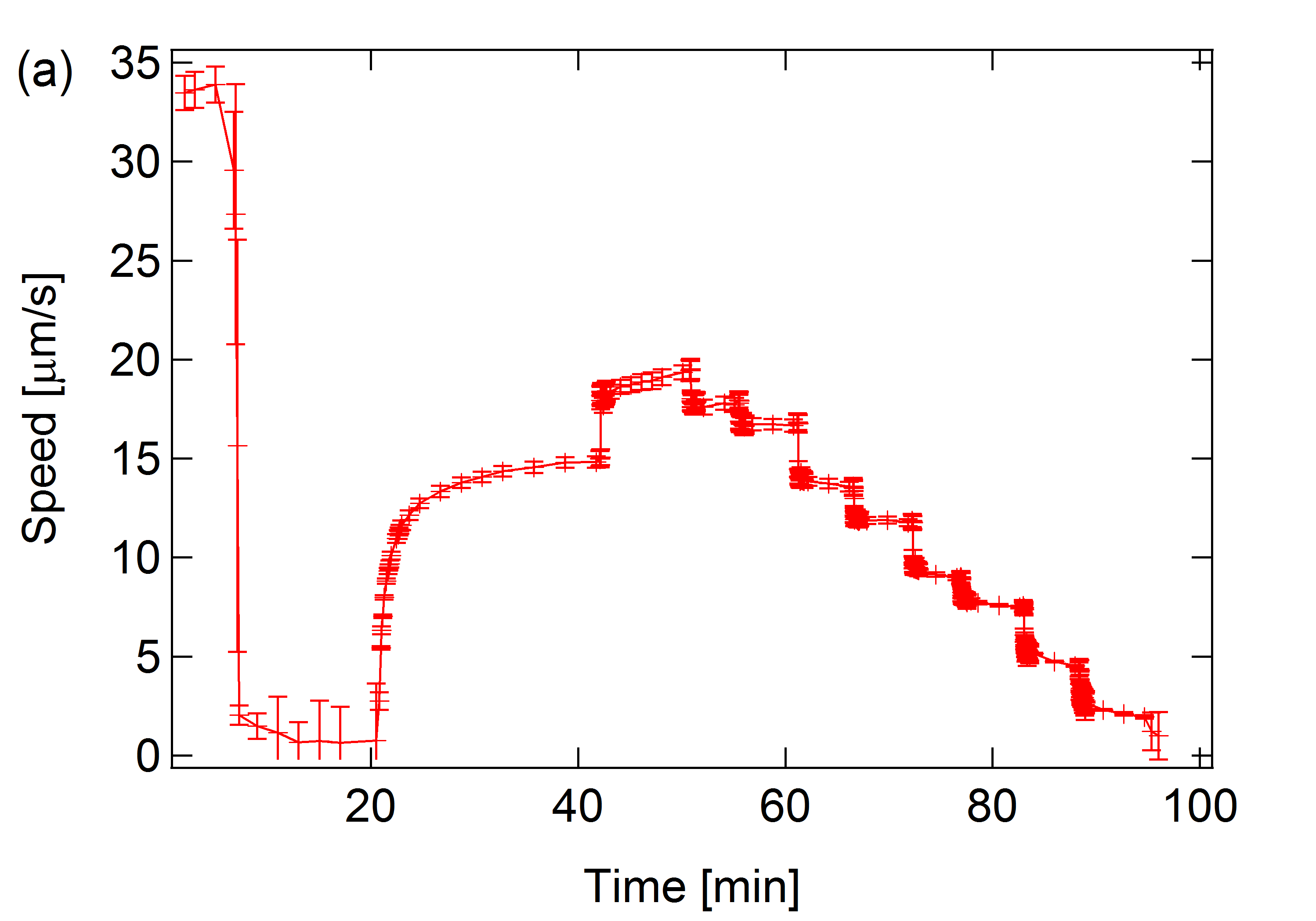}
	\includegraphics[width=0.48\textwidth]{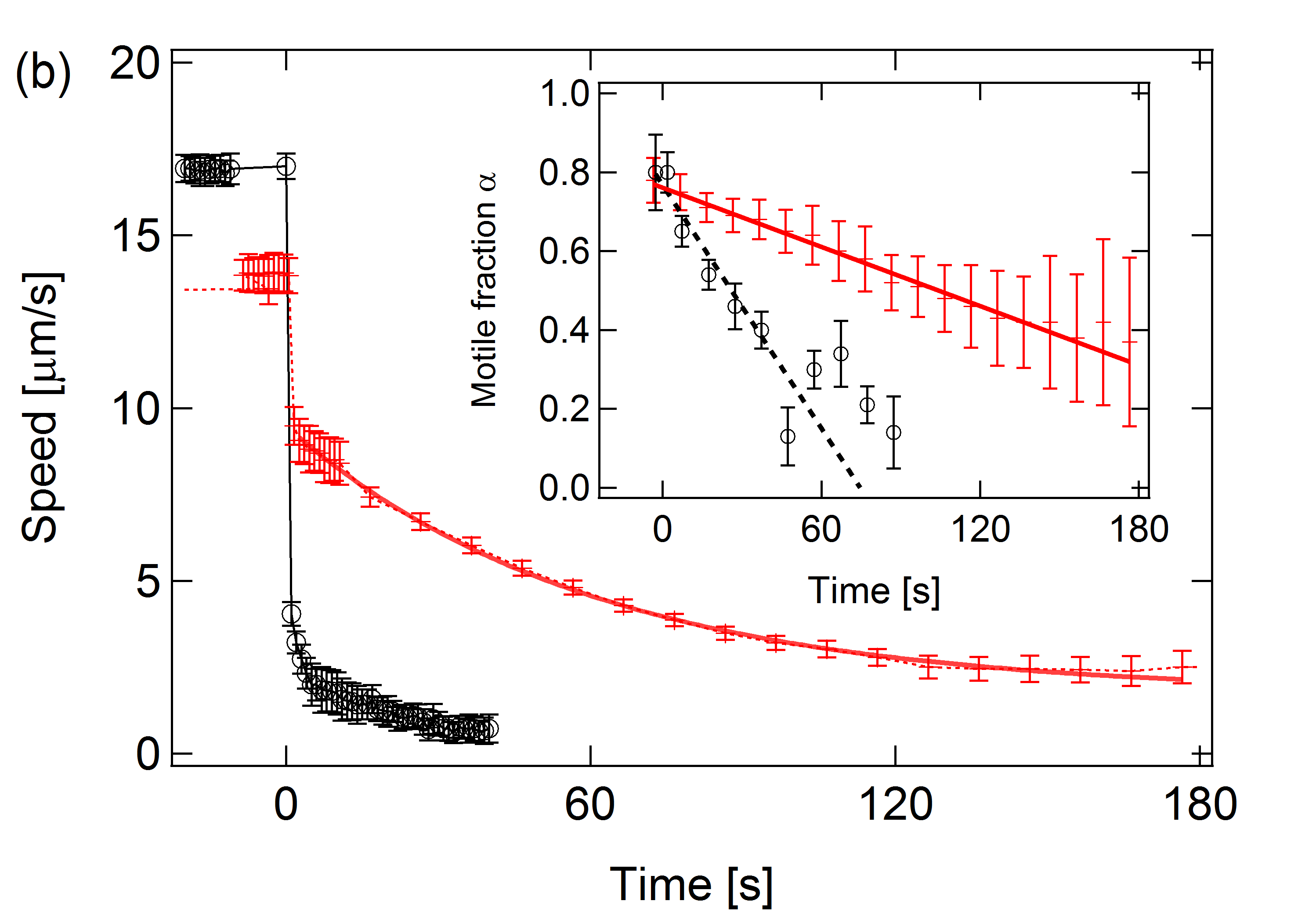}
	\caption{(a) Swimming speed of AD10 (sample OD = 1) during oxygen depletion and for different light intensities: Within $\approx \SI{7}{\minute}$ of sealing the sample oxygen was depleted and swimming ceased. Green light was switched on at $\approx \SI{20}{\minute}$. Illumination intensity was changed at various points throughout this experiment to provide data for the speed {\it vs.}~intensity plot in the main manuscript (inset, Fig.~\ref{fig:speed}). (b) A light-sensitive strain with {\it unc}-cluster deletion (AD10) shows a rapid decrease in swimming speed as soon as the illumination is stopped (black symbols). However, for strains which do not have the {\it unc} gene cluster deleted (AD57) a significant fraction of the population (see inset showing the motile fraction $\alpha=1-\beta$, lines are guides to the eye) continues to swim at reduced speed for several minutes, giving a much slower decay in $\bar v$ (red symbols; solid red line: exponential fit, giving $\tau \approx \SI{60}{\second}$).}
	\label{fig:SI:StartupAndIntensity}
\end{center}

\end{figure}

Figure~\ref{fig:speed} (inset) of the main text shows that $\bar v(\mathcal{I})$ rises linearly at low $\mathcal{I}$ and then saturates. The raw data for this plot are shown in Fig.~\ref{fig:SI:StartupAndIntensity}(a).

We claimed in the main text that deleting the {\it unc} gene cluster responsible for encoding the ATP synthase complex significantly reduced $\tau_{\rm off}$. Figure~\ref{fig:SI:StartupAndIntensity}(b) shows the behaviour of two strains of proteorhodopsin-powered {\it E. coli}: AD57 and AD10, without and with {\it unc} gene cluster deletion respectively. When illumination was switched off after the cells have reached steady state in constant illumination, the average swimming speed of both strains showed a sharp drop. In the case of AD10, this sharp drop was from $\sim 17$ to $\SI{4}{\micro\meter\per\second}$ over $\lesssim \SI{1}{\second}$. The drop was much smaller for AD57, from $\sim 14$ to $\SI{10}{\micro\meter\per\second}$, again over $\lesssim \SI{1}{\second}$. Thereafter, a significant fraction of AD57 cells kept on swimming with their average speed decreasing exponentially with a time constant of $\sim \SI{60}{\second}$.

\begin{figure}[thb]
\begin{center}
	\includegraphics[width=0.45\textwidth]{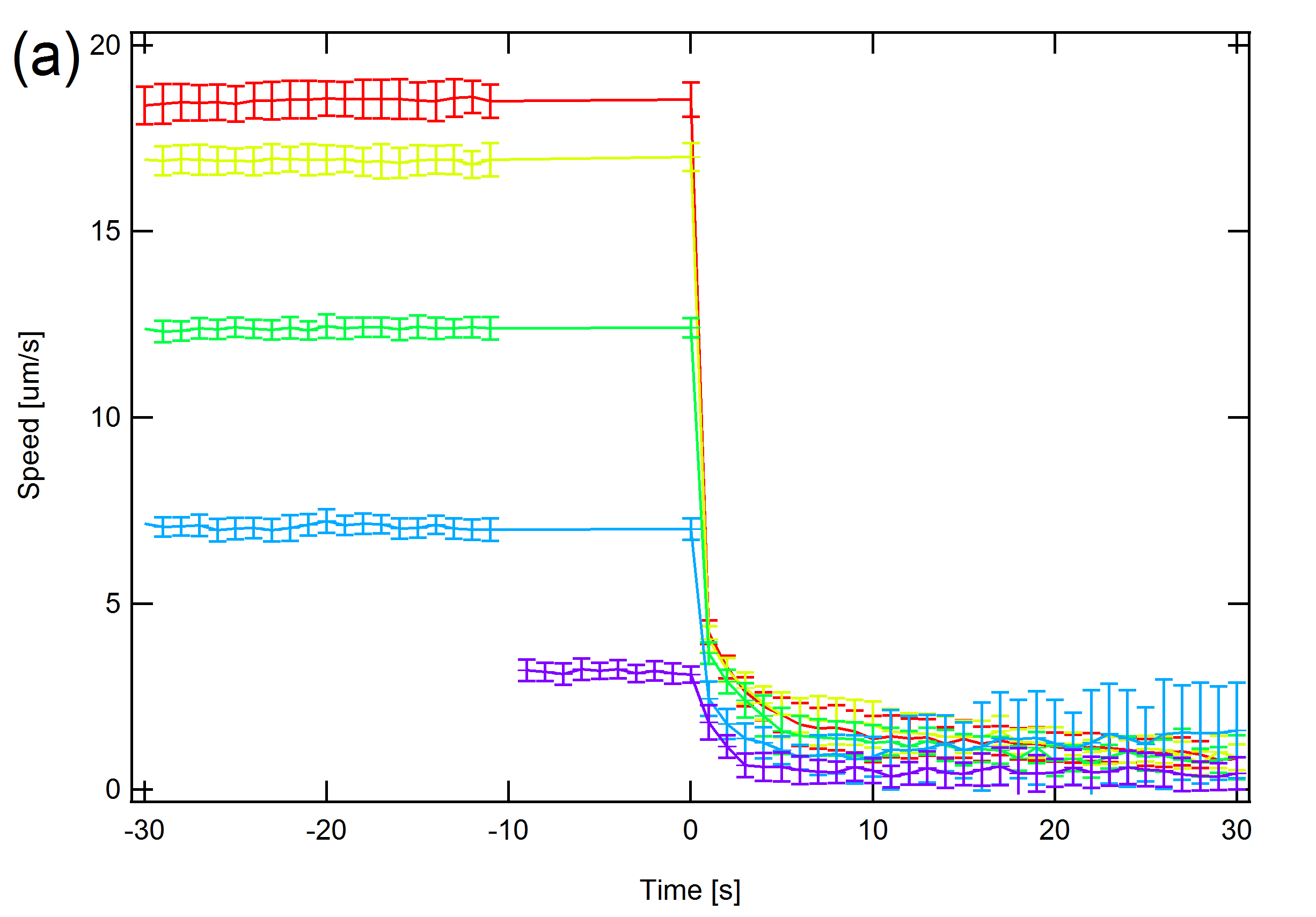}
	\includegraphics[width=0.45\textwidth]{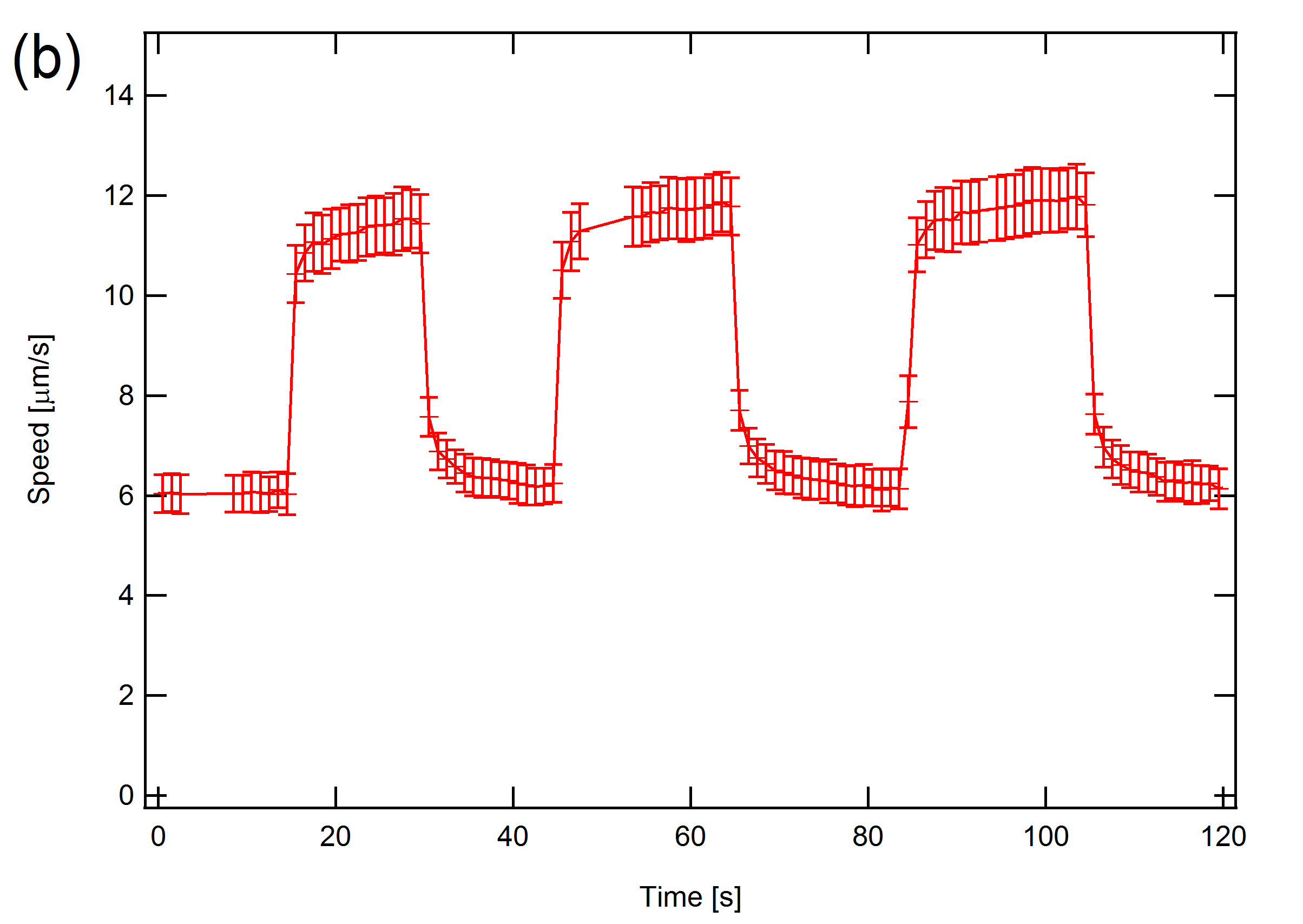}
	\caption{(a) Average swimming speed measured with DDM as green light is switched off (at $t=0$). There is a very rapid speed drop $\tau_{\rm off} \approx \SI{1}{\second}$ as the light is switched off, which is independent of the initial swimming speed (at the resolution limit of our technique). (b) AD10 bacteria adapt their swimming speed within 1s as the green light is alternated between low and high intensity. In this case $\tau_{\rm on} \approx \tau_{\rm off}$.}
	\label{fig:SI:StoppingVsSpeed}
\end{center}
\end{figure}

A simple resistor-capacitor equivalent circuit for the cell membrane and flagella motor was used in the main text to explain the observed $\tau_{\rm off} \approx \SI{1}{\second}$. We claimed that this was supported by the observation that $\tau_{\rm off}$ was independent of the starting speed from which cell decelerated. Figure~\ref{fig:SI:StoppingVsSpeed}(a) shows evidence for this claim. 

In the main text, we suggested that the long $\tau_{\rm on}$ of cells that have been kept in prolonged darkness was controlled by the time needed to reassemble the motor from individual stator units. We supported this claim by saying that $\tau_{\rm on}$ was much shorter for cells that had not been in low illumination for very long. Figure~\ref{fig:SI:StoppingVsSpeed}(b) shows evidence for this. When illumination alternated between high and low with a period of $\lesssim \SI{20}{\second}$, we find $\tau_{\rm} \lesssim \SI{1}{\second}$, comparable to $\tau_{\rm off}$. 

\subsection{Circuit model} \label{sec:SI3}

\begin{figure}[bht]
\begin{center}
	\includegraphics[width=0.3\textwidth]{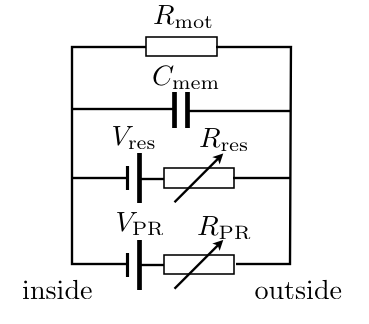}
	\caption{Equivalent circuit of the bacterial PMF generation and dissipation: $R_{\rm mot}$ = resistance of flagella motors, $C_{\rm mem}$ = membrane capacitance, $V_{\rm res}$ = PMF generated by respiratory enzymes, $R_{\rm res}$ = variable resistor modelling the effect of oxygen concentration on respiratory enzymes, $V_{\rm PR}$ = proteorhodopsin-generated PMF, $R_{\rm PR}$ = variable resistor modelling the effect of light on PR. (Redrawn after \cite{Walter2007}.)}
	\label{fig:SI:circuitmodel}
\end{center}
\end{figure}

The effect of proteorhodopsin (PR) on the proton motive force (PMF) can be modelled using a simple equivalent circuit, Fig.~\ref{fig:SI:circuitmodel}~\cite{Walter2007}. We use this model for our experiments by interpreting $R_{\rm res}$ as the effect of oxygen concentration on the respiratory circuit: $R_{\rm res} \rightarrow \infty$ when oxygen is exhausted. The effect of light on PR is modelled as a variable resistor, $R_{\rm PR}$, which Walters {\it et al.} have argued to be related to the incident light intensity $\mathcal{I}$ by $R_{\rm PR}(\mathcal{I}) = R_{\rm PR}^{\infty} (\mathcal{I} + \mathcal{K})/\mathcal{I}$ (with $\mathcal{K}$ a constant). Using the assumption that the average swimming speed is proportional to PMF, this leads to a similar intensity dependence of the speed $\bar v_{\rm sat}(\mathcal{I}) = \bar v_{\rm max} \mathcal{I}/(\mathcal{I}+\mathcal{I}_{1/2})$ under complete oxygen starvation.
Fitting this to our experimental data gives a maximum swimming speed $\bar v_{\rm max} = \SI{28.4 \pm 0.7}{\micro\meter\per\second}$ and an intensity to reach half this speed of $\mathcal{I}_{1/2} = \SI{10.7 \pm 0.4}{\milli\watt\per\centi\meter\squared}$, suggesting that our PR is significantly more efficient than the one used by Walter {\it et al.}~(for which $\mathcal{I}_{1/2} = \SI{60}{\milli\watt\per\centi\meter\squared}$) \cite{Walter2007}.

We suggest that the stopping time is controlled by the discharge of the PMF of the equivalent capacitance of the membrane through the equivalent resistance of the flagella motor, i.e.~$\tau_{\rm off} = R_{\rm mot}C_{\rm mem}$. The specific capacitance of bacterial membranes is $\approx \SI{0.6}{\micro\farad\per\square\centi\meter}$ \cite{Bai2006}. A $\SI{2}{\micro\meter} \times \SI{1}{\micro\meter}$ spherocylindrical {\it E. coli} cell has $2\pi \,\si{\square\micro\meter}$ surface area, giving $C_{\rm mem} \approx 4 \times 10^{-14} \si{\farad}$. We have previously estimated \cite{Poon2016a} that the proton current through a motor is $\approx 2 \times 10^4$ \ce{H+} \si{\per\second}, which is driven by a PMF of $\approx \SI{-150}{\milli\volt}$, giving $ R_{\rm mot} \approx 0.5 \times 10^{14} \si{\ohm}$ per motor. On average, each cell has, say, $\approx 5$ motors. These constitute resistors in parallel, so that $R_{\rm mot} \approx 0.1 \times 10^{14} \si{\ohm}$. Thus, $\tau_{\rm off} \sim R_{\rm mot}C_{\rm mem} \lesssim \SI{1}{\second}$, as claimed in the main text.

\subsection{Additional data for cells leaving illuminated squares} \label{sec:SI:squares}

\begin{figure}[t]
\begin{center}
\includegraphics[width=0.95\columnwidth]{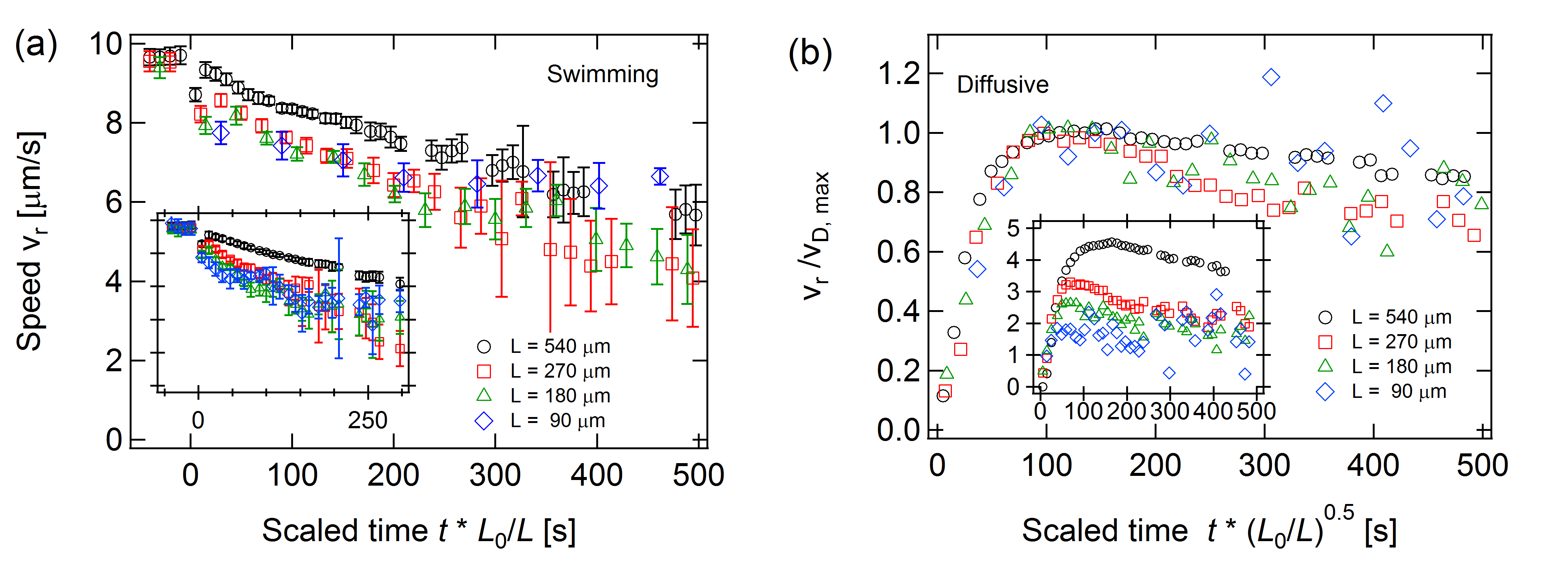}
\caption{Average swimming speed inside emptying squares of various sizes $L$ when starting with swimming or diffusive cells. For swimming cells (a) the speed does not quite stay constant at $v_{\rm sat} = \SI{9.5}{\micro\meter\per\second}$ but drops about linearly with time.  For diffusive cells (b) the speed displays more complex dynamics: the swimming speed $v$ initially increases linearly with time before approaching a maximum value which increases with the size of the square $L$. Scaling speed with $v_{\rm max}$ and time with $\sqrt{L/L_0}$ leads to good data collapse (with $L_0=540\mu m$).  (Insets: raw speed data against time without any scaling).
}
\label{fig:SI:SpeedWithSize}
\end{center}
\end{figure}

In the main text we reported measurements of how long it took to empty an illuminated square (side $L$) of cells, starting from swimming or stationary cells. The emptying time scales as $L$ and $L^{1/2}$ respectively. 
This is the scaling expected for random ballistic swimmers leaving a box of size $L$. This is appropriate for our experiments, as the persistence length of our smooth swimming bacteria is comparable to the box size. 
Here we report the average speed of cells remaining in the illuminated square, $\bar v_{\rm r}$, as a function of time in the two cases. 
Starting from swimming cells, our simplistic model presented in the main manuscript assumed a constant swimmming speed $\bar v_{\rm r}=\bar v_{\rm sat}$, whereas the actually measured speed $\bar v_{\rm r}$ decreased with time, Fig.~\ref{fig:SI:SpeedWithSize}(a): our bacteria have a broad swimming speed distribution and the fastest-swimming cells leave the square first, leading to a change in $P(v)$ and a corresponding drop in the mean speed. 
This drop becomes more rapid the smaller the square, displaying approximately the same scaling with $L$ as the average cell density shown in the main text. 

Starting from stationary (meaning diffusing) cells, $\bar v_{\rm r}(t)$ is non-monotonic, Fig.~\ref{fig:SI:SpeedWithSize}(b). Immediately after illumination inside the square is switched on, cells accelerate over a time scale of $\sim \tau_{\rm on}$; later, $\bar v_{\rm r}$ decreases with time for the same reason as in the case of starting with swimming cells: the fastest cells leave earlier. 
The maximum speed reached by a swimmer depends on the time it spent in the box, leading to $v_{\rm D, max} = a \mathfrak{t}_{\rm D} \sim \sqrt{L}$, based on $\mathfrak{t}_{\rm D} \sim \sqrt{L}$ as demonstrated in the main text. This is in agreement with the measured population averaged $\bar v_{\rm D, max} \sim \sqrt{L}$ shown in Fig.~\ref{fig:scaling}c of the main manuscript, and the data collapse of  $\bar v_{\rm r}/\bar v_{\rm D, max}$ with scaled time $t(L_0/L)^{0.5}$ as shown in Fig.~\ref{fig:SI:SpeedWithSize}(b).

\subsection{Differential dynamic microscopy} \label{sec:SI:DDM}

\begin{figure}[t]
\begin{center}
\includegraphics[width=.5\columnwidth]{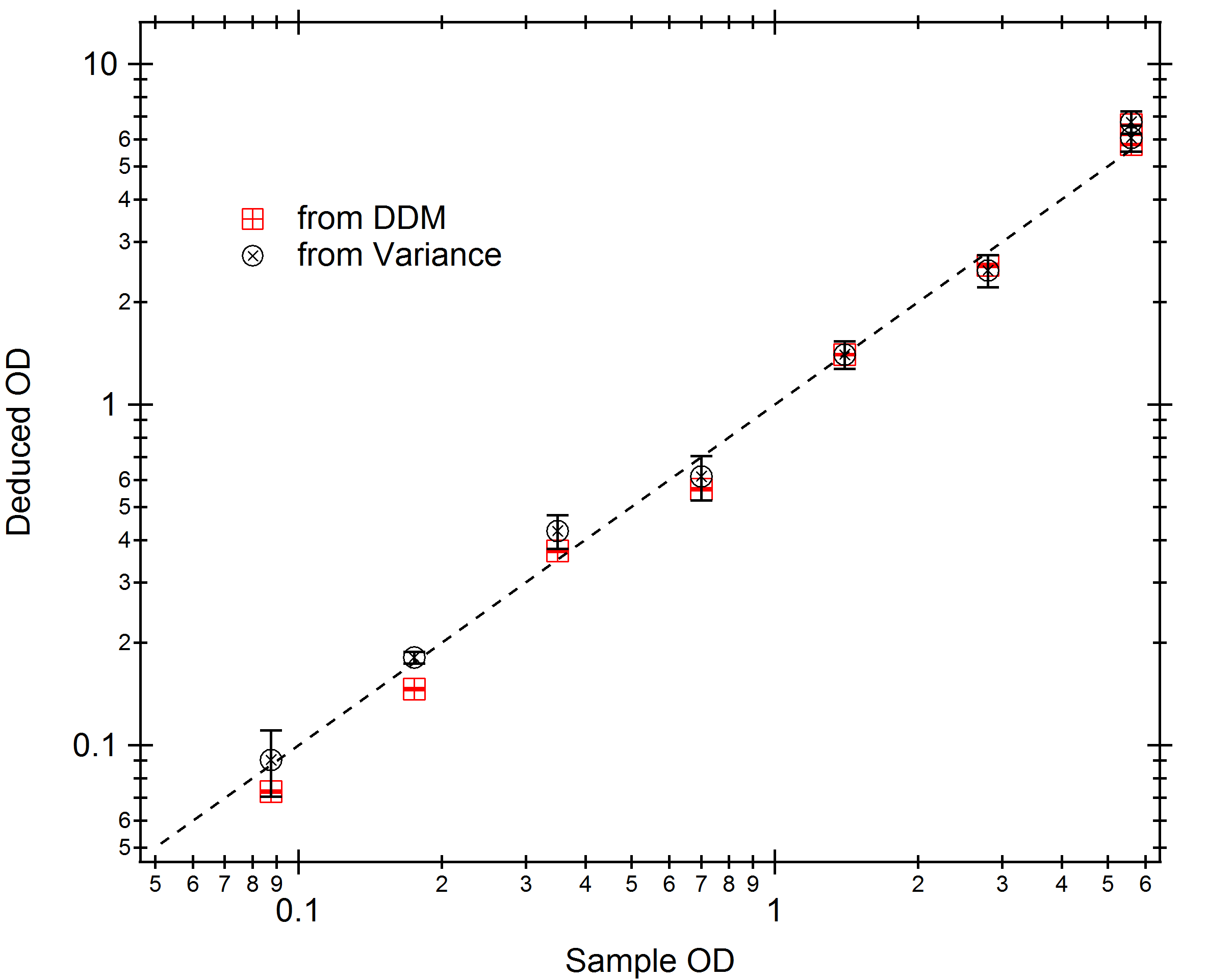}
\caption{Cell density deduced from DDM (\S\ref{sec:SI:DDM}) and intensity variance (section \S\ref{sec:SI:Variance}) measurements for a series of 7 samples starting at a measured optical density of OD = 5.6 and diluted successively by factors of 2. Both analysis methods were applied to the same movies and yield consistent results in quantitative agreement with the prepared sample densities throughout the measured range. Both data sets have been calibrated by assuming the data point for OD 1.4 was measured correctly.}
\label{fig:SI:DensityDDM-Var}
\end{center}
\end{figure}

Differential dynamic microscopy (DDM) is a fast, high-throughput technique to measure the dynamics of particle suspensions such as colloids \cite{CerbinoDDM1} or swimming microorganisms \cite{WilsonDDM, MartinezDDM, Poon2016a}. For swimming bacteria, DDM measures   motility parameters averaged over $\sim10^4$ cells from $\approx \SI{30}{\second}$ long movies via the differential image correlation function $g(\vec{q},\tau)$, i.e.~the power spectrum of the difference between pairs of images delayed by time $\tau$. Under appropriate imaging conditions and for isotropic motion, $g(\vec{q},\tau)$ is related to the intermediate scattering function $f(q, \tau)$, which is the $q^{\rm th}$ Fourier component of the density temporal autocorrelation function, via
\begin{equation}
  g(q,\tau) = A(q)\left[1 - f(q,\tau)\right] + B(q) \,.
\end{equation} 
Here, $B(q)$ relates to the background noise and $A(q)$ is the signal amplitude. Fitting $f(q,\tau)$ to a suitable swimming model of {\it E. coli}, we obtain the average, $\bar{v}$, and width, $\sigma$, of the speed distribution $P(v)$, the fraction of non-motile bacteria $\beta$ and their diffusion coefficient $D$ \cite{MartinezDDM}. Provided the system is dilute enough such that its structure factor $S(q) \approx 1$, $A(q)$ is proportional to the particle density \cite{Reufer,ConfocalDDM}. Thus, in two otherwise identical samples with different cell densities $\rho_{1,2}$, 
\begin{equation}
  \frac{A_1(q)}{A_2(q)} = \frac{\rho_1}{\rho_2}\,. \label{eq:Aq}
\end{equation}
We verified this using bacteria suspensions of different cell concentrations (measured by optical density), Fig.~\ref{fig:SI:DensityDDM-Var}.

\subsection{Estimating cell density using intensity variance} \label{sec:SI:Variance}

\begin{figure}[t]
\begin{center}
\includegraphics[width=0.9\columnwidth]{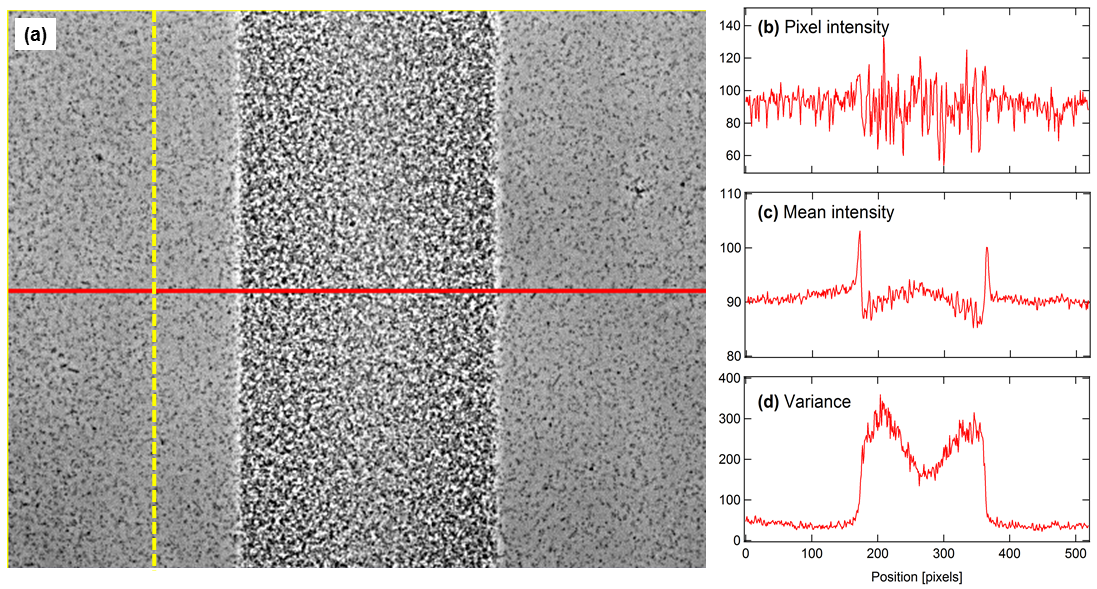}
\caption{Extracting density profiles from phase contrast images: (a) phase contrast image with a line (red) used for the pixel intensity profile, one vertical line (dashed yellow) used for calculating the mean intensity and variance at this horizontal position. Intensity profile of (b) single pixel line and the (c) averaged profile, which is dominated by the bright halo artifacts next to the edges. (d) Intensity variance profile.}
\label{fig:SI:variance}
\end{center}
%
%
\end{figure}

Although DDM in the form of Eq.~\ref{eq:Aq} is useful for extracting average cell densities over large uniform areas, its strictly limited spatial resolutions makes it unsuitable for assessing cell densities that vary rapidly in space, such as from images shown in Fig.~\ref{fig:patterns} of the main text to illustrate STASA, or the band of high density bacteria shown in Fig.~\ref{fig:SI:variance}(a). In the latter case, neither the intensity along a line nor the average of such line profiles, Fig.~\ref{fig:SI:variance}(b) and (c) respectively, bears much resemblance to the cell density variation visible to direct inspection. This is related to the nonlinearities at high phase shifts and artifacts inherent to phase contrast imaging~\cite{Pluta89b}.

Interestingly, the single line profile, Fig.~\ref{fig:SI:variance}(b), shows that the intensity fluctuates more in the high-density region.
Indeed, the averaged horizontal variance profile, Fig.~\ref{fig:SI:variance}(d), shows qualitative agreement with the visually-observed density profile of the band, suggesting that the variance (with background subtracted) could be used to quantify local density. Figure \ref{fig:SI:DensityDDM-Var} shows that, indeed, densities of uniform samples deduced from variance measurements agree quantitatively with densities deduced from DDM using Eq.~\ref{eq:Aq} and from optical density measurements.
At least within the low density limit, this scaling of intensity variance with cell density is not surprising: a (dark) cell on a uniform (light) background will lower the mean intensity slightly and introduce a characteristic amount of variance relative to this background. As more cells are added they initially increase the variance by the same characteristic amount (as the mean background hardly changes), leading to the observed proportionality. For phase contrast imaging, phase-dark object are often surrounded by a lighter halo, so that the net effect of cells on mean intensity is rather small while at the same time leading to a significant characteristic variance per object.

\subsection{STASA for AD10 vs AD57} \label{sec:SI:AD10vsAD57}

We demonstrate the effect of $\tau_{\rm stop}$ on STASA by projecting a bright-dark intensity step onto cells that had been uniformly illuminated for $\approx\SI{10}{\minute}$ (so that the population had reached saturation speed, $\bar v_{\rm AD10} = \SI{17.5}{\micro\meter\per\second}$ and $\bar v_{\rm AD57} = \SI{11.8}{\micro\meter\per\second}$).

\begin{figure}[hp]
\begin{center}
\includegraphics[width=1\columnwidth]{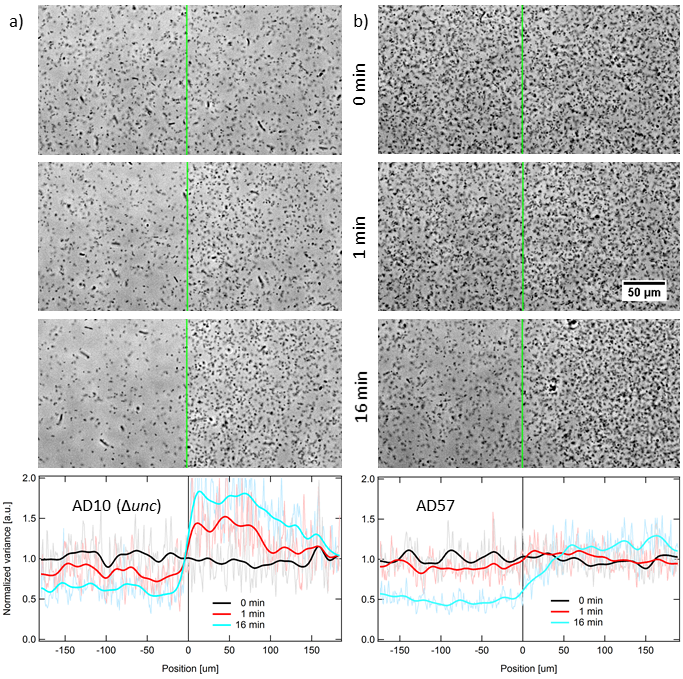}
\caption{Spatial control of a) AD10 strain compared to b) AD57 strain. Phase contrast images (Nikon PF 20x/0.5 objective) of the samples just before ($\SI{0}{\minute}$) as well as  $\SI{1}{\minute}$ and $\SI{16}{\minute}$ after blocking the green light on the right half of the image (boundary position indicated by green lines). 
The corresponding variance profiles (normalised by the mean variance under uniform illumination) are shown underneath the images. 
The numerically smoothing (bold lines; using a binomial smoothing filter with $n=201$) of the raw data (faint lines) highlights the underlying change in cell density. For the AD57 strain the density difference develops much more slowly and shows a much more gradual transition, making it less suitable for STASA.}
\label{fig:SI:edge}
\end{center}
\end{figure}

Fig. \ref{fig:SI:edge} shows snapshots of samples of a) the fast-stopping AD10 strain and b) the AD57 strain just before ($\SI{0}{\minute}$) as well as  $\SI{1}{\minute}$ and $\SI{16}{\minute}$ after introducing the intensity step. For AD10 a clear difference in cell density between two sharply defined halves is apparent very quickly, with cells entering the dark region stopping almost instantaneously. Cells in the illuminated half keep on swimming until they enter a dark region, leading to a drop in cell number in the illuminated region and an accumulation of cells just inside the dark region.
For AD57 the cells in the dark half initially only slow down but keep on swimming at a lower speed for several minutes. Therefore there is initially very little effect on the cell density, and only after a much longer time a far more gradual pattern emerges, as demonstrated by the profiles of the intensity variance (reproduced from the main text in Fig.~\ref{fig:SI:edge}).

\clearpage
\subsection{Movies}

\setcounter{figure}{0} 
\renewcommand{\figurename}{Movie}
\renewcommand{\thefigure}{M\arabic{figure}}%

\begin{figure}[hb]
\begin{center}
\href{run:./SI-Movie1-UoE-startup-2fps-5xSpeedup-Time-uncomp.avi}{\includegraphics[width=.45\columnwidth]{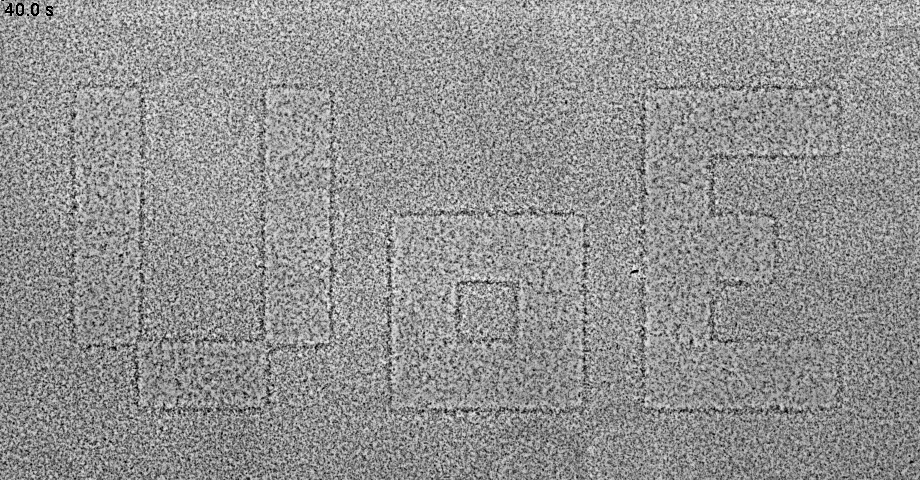}}
\caption{$\SI{40}{\second}$ movie showing the initial formation of the `UoE' pattern. This corresponds to the run--up to fig.~\ref{fig:patterns}a in the main manuscript. The frame--rate has been reduced from the 100\,fps of the original movie to 2\,fps and is played back at $5\times$ the real speed. One pixel corresponds to $\SI{1.4}{\micro\meter}$.  }
\label{mov:SI:startup}

\end{center}
\end{figure}

\begin{figure}[hb]
\begin{center}
\href{run:./SI-Movie2-UoE-inversion-2fps-5xSpeedup-Time-uncomp.avi}{\includegraphics[width=.45\columnwidth]{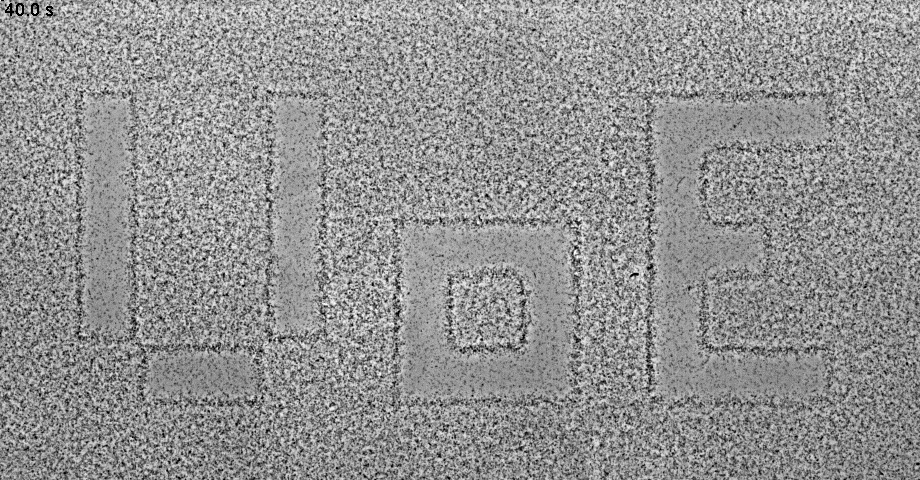}}
\caption{Movie showing initial  $\SI{40}{\second}$ just after the inversion of the `UoE' pattern. This corresponds to the transition between fig.~\ref{fig:patterns}c and d in the main manuscript. Frame--rate adjusted as for movie M1.}
\label{mov:SI:inversion}
\end{center}
\end{figure}

\begin{figure}[hb]
\begin{center}
\href{run:./SI-Movie3-UoE-TimeLapse.avi}{\includegraphics[width=.45\columnwidth]{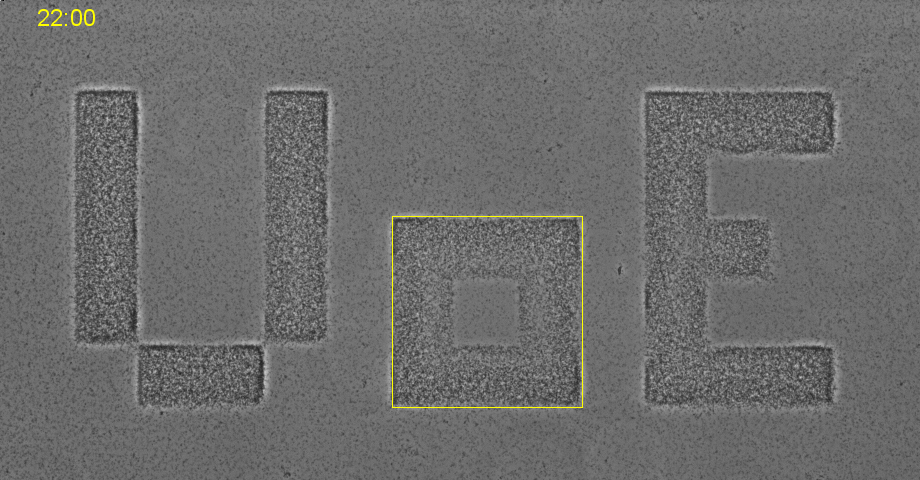}}
\caption{Timelapse image series of the UoE pattern being formed and inverted. This illumination pattern is switched on at time `0:00' when the sample is still uniform (corresponding to movie~M1) and becomes clearer over time. After $\SI{9}{\minute}$ the pattern is inverted (corresponding to movie~M2) and again becomes more pronounced over time. The outline of the letter `o' is marked to highlight the switching of the boundary as the pattern is inverted.
}
\label{mov:SI:timelapse}

\end{center}
\end{figure}

\clearpage

\bibliography{active}

\end{document}